# Strong Fault-Tolerance for Self-Assembly with Fuzzy Temperature

David Doty* Matthew J. Patitz† Dustin Reishus‡ Robert T. Schweller§
Scott M. Summers¶

**Abstract**

We consider the problem of fault-tolerance in nanoscale algorithmic self-assembly. We employ a standard variant of Winfree's abstract Tile Assembly Model (aTAM), the *two-handed* aTAM, in which square "tiles" – a model of molecules constructed from DNA for the purpose of engineering self-assembled nanostructures – aggregate according to specific binding sites of varying strengths, and in which large aggregations of tiles may attach to each other, in contrast to the *seeded* aTAM, in which tiles aggregate one at a time to a single specially-designated "seed" assembly. We focus on a major cause of errors in tile-based self-assembly: that of unintended growth due to "weak" strength-1 bonds, which if allowed to persist, may be stabilized by subsequent attachment of neighboring tiles in the sense that at least energy 2 is now required to break apart the resulting assembly; i.e., the errant assembly is *stable at temperature 2*.

We study a common self-assembly benchmark problem, that of assembling an $n \times n$ square using $O(\log n)$ unique tile types, under the two-handed model of self-assembly. Our main result achieves a much stronger notion of fault-tolerance than those achieved previously. *Arbitrary* strength-1 growth is allowed; however, any assembly that grows sufficiently to become stable at temperature 2 is guaranteed to assemble into the correct final assembly of an $n \times n$ square. In other words, errors due to insufficient attachment, which is the cause of errors studied in earlier papers on fault-tolerance, are prevented *absolutely* in our main construction, rather than only with high probability and for sufficiently small structures, as in previous fault-tolerance studies. We term this the *fuzzy temperature* model of faults, due to the following equivalent characterization: the temperature is normally 2, but may drift down to 1, allowing unintended temperature-1 growth for an arbitrary period of time. Our construction ensures that this unintended growth cannot lead to permanent errors, so long as the temperature is eventually raised back to 2. Thus, our construction overcomes a major cause of errors, insufficient strength-1 attachments becoming stabilized by subsequent growth, without requiring the detachment of strength-2 bonds that slows down previous constructions, and without requiring the careful fine-tuning of thermodynamic parameters to balance forward and reverse rates of reaction necessary in earlier work on fault-tolerance.

Although we focus on the task of assembling an $n \times n$ square, our construction uses a number of geometric motifs and synchronization primitives that will likely prove useful in other theoretical (and, we hope, experimental) applications.

*Univ. of Western Ontario, Dept. of Computer Science, London, Ontario, Canada, N6A 5B7, `ddoty@csd.uwo.ca`.
This author was supported in part by National Science Foundation Grants 0652569 and 0728806.

†Univ. of Texas–Pan American, Dept. of Computer Science, Edinburg, TX, USA, 78539, `mpatitz@cs.panam.edu`.
This author was supported in part by National Science Foundation Grants 0652569 and 0728806.

‡Univ. of Southern California, Dept. of Computer Science, Los Angeles, CA, USA, 90089, `reishus@usc.edu`

§Univ. of Texas–Pan American, Dept. of Computer Science, Edinburg, TX, USA, 78539, `schwellerr@cs.panam.edu`

¶Iowa State Univ., Dept. of Computer Science, Ames, IA, USA 50011, `summers@cs.iastate.edu`. This author was supported in part by National Science Foundation Grants 0652569 and 0728806.

# 1 Introduction

Tile-based self-assembly is a model of "algorithmic crystal growth" in which square "tiles" represent molecules that bind to each other via specific and variable-strength bonds on their four sides, driven by random mixing in solution but constrained by the local binding rules of the tile bonds. Beginning with the experimental work of Seeman in the early 1980s [38], such molecules have been engineered from DNA in the laboratory, and used to create a variety of sophisticated self-assembled structures [7, 13, 26, 29, 30, 35, 47] such as Sierpinski triangles and binary counters. Erik Winfree [43,44], based on experimental work of Seeman [38], modified Wang's mathematical model of tiling [41,42] to add a physically plausible mechanism for growth through time. Winfree defined two models of tile-based self-assembly, the abstract Tile Assembly Model (aTAM) and the kinetic Tile Assembly Model (kTAM). In both models, the fundamental components are un-rotatable, but translatable square "tile types" whose sides are labeled with glue "colors" and "strengths." Two tiles that are placed next to each other *interact* if the glue colors on their abutting sides match, and in the aTAM, a tile *binds* to an assembly if it interacts on all sides with total strength at least a certain ambient "temperature," usually taken to be 2. In particular, if a tile has two strength-1 glues, both of them must match the corresponding glues in the assembly in order to remain bound.

In the more thermodynamically plausible kTAM, tiles may bind even if they interact with strength less than 2, but are assumed to detach at a rate inversely and exponentially proportional to the strength with which they interact. Hence tiles attached with strength 1 detach "quickly", and tiles attached with strength 2 detach "slowly". A tile attached with only strength 1 (a so-called "insufficient attachment") represents a potential error, as its other strength-1 glue may be mismatched with the abutting portion of the assembly, or mismatched with what is eventually intended to be placed at that position. However, since strength-1 attachments are assumed to detach after a short time, an insufficient attachment actualizes into a permanent error only if another tile first binds to secure the faulty tile in place, causing the entire assembly to become stable at temperature 2. That is, by "wandering" temporarily through the space of assemblies producible at temperature 1, we may arrive at an assembly not producible at temperature 2, yet that, once formed, is stable at temperature 2. The development of physical and algorithmic mechanisms for preventing such errors remains a formidable challenge in nanoscale self-assembly.

Stated informally, the kTAM refines the aTAM by endowing it with a mechanism for error (temporary binding of tiles with strength 1) as well as a mechanism for error correction (eventual detachment of tiles, even those bound with strength 2). Indeed, numerous papers have used these two mechanisms for high-probability error correction in the kTAM [11–13, 31, 44, 46]. In each of these papers except [44], the same basic principle is used to achieve error correction, known as *proofreading*. If an insufficient attachment results in mismatching glues, this error is "amplified" by forcing further growth to require many other insufficient attachments to stabilize. Since these happen only slowly, the assembly process is slowed down, giving time for the tiles that stabilized the original insufficient attachment to detach, thus correcting the original error. In [44], Winfree also shows how errors are removed through the detaching of tiles, although there is no "error-amplification process"; Winfree shows that by setting the ratio of the forward rate to the reverse rate sufficiently small (thus slowing down the entire assembly process), erroneous tiles will detach with high probability. Other papers have investigated algorithmic correction of other types of errors [27,36,37,39,45] and physical, rather than algorithmic, mechanisms for error suppression [10,17,28].

We work in a variant of the aTAM known as the *two-handed* aTAM [1,2,4,6,14,27,45]. Winfree's original model, the *seeded* aTAM [43,44], stipulates that assembly begins from a specially-designated

"seed" tile type, and all binding events consist of the attachment of a single tile to the growing assembly that contains the seed. The seed thus serves as a *nucleation point* from which all further growth occurs. In reality, such single-point nucleation is difficult to enforce [36, 37] as tiles with matching glues may attach to each other in solution, even if neither of them is connected to the seed tile. The two-handed aTAM models this sort of growth by dispensing with the idea of a seed, and simply defining an assembly to be producible if 1) it consists of a single tile (base case), or 2) it results from the stable aggregation of two producible assemblies (recursive case). We emphasize that the present paper does not introduce this two-handed version of the model; it has been studied by numerous authors [1, 2, 4, 6, 14, 27, 45] under various names, such as the "multiple tile" model or the "polyomino" model.

Not only is the two-handed aTAM a more realistic model in the sense of accounting for unseeded nucleation, it allows us to use the geometry of partially-formed assemblies, rather than relying solely on (error-prone) glue specificity, to enforce binding rules between subassemblies. This phenomenon, geometric blocking that prevents bond formation, is well-studied in chemistry and is known as *steric hindrance* [40, Section 5.11] or, particularly when employed as a design tool for intentional prevention of unwanted binding in synthesized molecules, *steric protection* [19–21]. Using the mechanism of steric protection, we are able to achieve a much stronger notion of fault-tolerance than that described in previous error-correction papers. Informally, our model of fault-tolerance, which we term the *fuzzy temperature* model, is as follows (a formal description is given in Section 4). Similarly to the kTAM, we allow strength-1 insufficient attachments to occur. However, we do not model forward or reverse rates of growth as in the kTAM, as there is no need to employ the higher reverse rates of insufficient attachments: any insufficient attachments that lead to an assembly that is stable at temperature 2 *were never errors in the first place*, as such an assembly can always lead to an assembly that was producible with only strength-2 growth. That is, viewed as a modification of the aTAM, we allow the temperature to be "fuzzy", occasionally drifting from 2 down to 1, which allows strength-1 growth for as long as the temperature remains low. However, once the temperature is raised back to 2, thus dissolving any structure that is stable only at temperature 1, the stable assemblies that are left over are all assemblies that are already producible at temperature 2 or that can grow into a temperature-2-producible assembly. Therefore, while insufficient attachments can occur, errors due to insufficient attachments cannot occur, since temperature-2 stabilization of such errors, which our construction prevents, is required for the errors to become permanent.

We focus on the problem of assembling an $n \times n$ square, a common benchmark problem for demonstrating the use of self-assembly techniques [3, 6, 8, 15, 22, 23, 34]. In particular, our main result is the construction of a tile set with $O(\log n)$ unique tile types (which is close to the $\Omega(\log n / \log \log n)$ optimal lower bound [34]) that uniquely assembles into an $n \times n$ square in the two-handed aTAM at temperature 2, and that has the fuzzy-temperature fault-tolerance property described above. In keeping with the "wandering" analogy from the beginning of this section, our construction allows arbitrary wandering in the space of assemblies producible at temperature 1, but funnels all such wandering towards a single unique terminal assembly, or towards the oblivion of destruction at temperature 2.[1]

[1] We emphasize that this is *not* the same as saying that our construction assembles an $n \times n$ square at temperature 1: at temperature 1, many different terminal assemblies can nondeterministically form, most of which are junk. Our construction ensures that when the temperature is raised to 2, all the junk dissolves away, leaving only assemblies that are required to assemble the square, and which could have grown anyway had the temperature remained at 2. In fact, it is an open problem, first stated by Winfree and Rothemund in [34], to uniquely assemble an $n \times n$ square

This paper is organized as follows. Section 2 gives an informal description of the two-handed aTAM. Section A formally defines the two-handed aTAM. Section 3 shows a construction of a non-fault-tolerant counter, to introduce some of the main ideas of the full construction. Section 4 defines the fuzzy temperature model of fault-tolerance. Section 5 describes a high-level overview of the main construction and explains the basic techniques employed. Section B explains the main construction in detail. Section 6 concludes the paper and states open questions.

# 2 Informal Description of the Two-Handed Abstract Tile Assembly Model

This section gives a brief informal sketch of the two-handed temperature-2 abstract Tile Assembly Model (aTAM). The model is described formally, and more generally, in Section A.

A *tile type* is a unit square with four sides, each having a *glue* consisting of a *label* (a finite string) and *strength* (0, 1, or 2). We assume a finite set $T$ of tile types, but an infinite number of copies of each tile type, each copy referred to as a *tile*. A *supertile* (a.k.a., *assembly*) is a positioning of tiles on the integer lattice $\mathbb{Z}^2$. Two adjacent tiles in a supertile *interact* if the glues on their abutting sides are equal and have positive strength. Each supertile induces a *binding graph*, a grid graph whose vertices are tiles, with an edge between two tiles if they interact. The supertile is $\tau$*-stable* if every cut of its binding graph has strength at least $\tau$, where the weight of an edge is the strength of the glue it represents. That is, the supertile is stable if at least energy $\tau$ is required to separate the supertile into two parts. A *tile assembly system* (TAS) is a pair $\mathcal{T} = (T, \tau)$, where $T$ is a finite tile set and $\tau$ is the *temperature*, usually 1 or 2. Given a TAS $\mathcal{T} = (T, \tau)$, a supertile is *producible* if either it is a single tile from $T$, or it is the $\tau$-stable result of translating two producible assemblies. A supertile $\alpha$ is *terminal* if for every producible supertile $\beta$, $\alpha$ and $\beta$ cannot be $\tau$-stably attached. A TAS is *directed* (a.k.a., *deterministic*, *confluent*) if it has only one terminal, producible supertile. Given a connected shape $X \subseteq \mathbb{Z}^2$, a TAS $\mathcal{T}$ *produces* $X$ *uniquely* if every producible, terminal supertile places tiles only on positions in $X$ (appropriately translated if necessary).

# 3 Two-Handed Assembly of a Counter from $O(\log n)$ Tile Types

In this section we describe the two-handed assembly of a (non-fault-tolerant) counter from $O(\log n)$ tile types, as a warmup to our full fault-tolerant square construction. While this technique does not achieve fault-tolerance, it introduces a novel new counter design technique that utilizes 1) geometry to enforce/restrict specific assemblies and 2) non-determinism of supertile formation and attachment to explore the space of possible intermediate assemblies, despite the existence of only one unique terminal assembly. This technique forms the basis for the more involved fuzzy fault tolerant construction.

The tile set for the counter is depicted in Figure 1. In this figure, an example tile set for a 4 bit counter that counts from 0 to 15 is provided. Tile types that share unique, full strength $\tau = 2$ glues are connected by a black line that crosses over the bonded edge. Other glues in the system include strength $\tau = 2$ glues $A_i$ and $C_i$ for $i$ from 0 to $\log n$ for a length $n$ counter, and two strength $\tau = 1$ glues denoted by the green and blue squares.

at temperature 1 using fewer than $2n - 1$ tile types (compared to our use of $O(\log n)$ tile types). Rothemund and Winfree conjectured that $2n - 1$ is a strict lower bound for this problem.

Conceptually, the tile set of Figure 1 consists of a number of blocks for each bit position of a binary counter. These blocks assemble into height $O(\log n)$ columns, where the representative block for each bit is determined non-deterministically. Further, the geometry of each block encodes a bit on both the left and right side of the block by a *dent* that appears at either the upper or lower half of the block. In the case of orange *rollover* blocks, the left side encodes the value 1, while the right encodes the value 0; these represent 1 bits less significant than the least significant 0, which all change from 1 to 0 on the next increment. For the yellow *least significant 0* blocks, the left dent encodes the value 0 and the right encodes 1. For the grey *copy* blocks, the left and right encode the same value, with one type of grey block for "1" and another for "0"; these represent bits more significant than the least significant 0, which remain the same on the next increment.

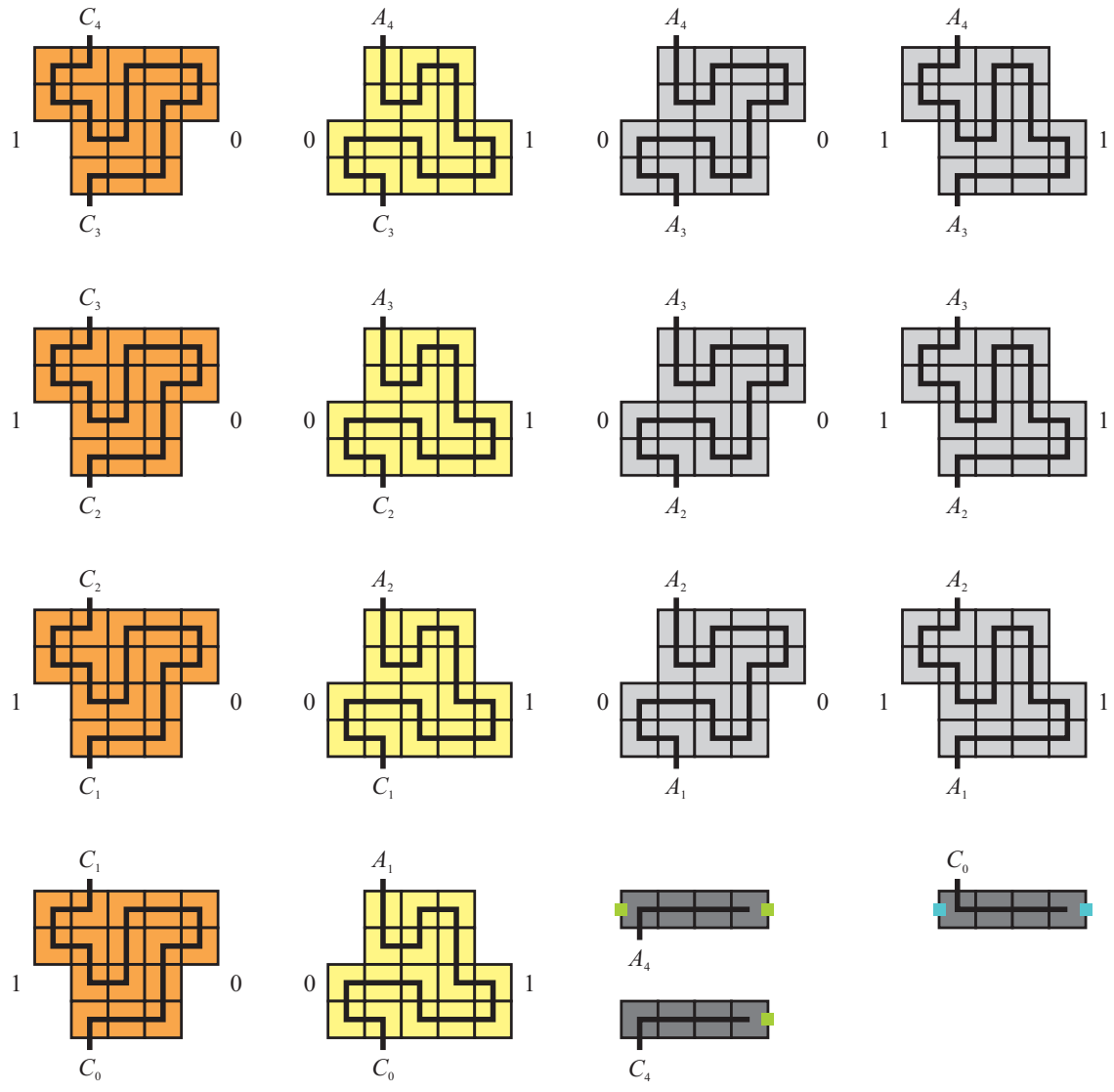

Figure 1: Tile set for two-handed assembly of a length $n$ binary counter using $O(\log n)$ tile types.

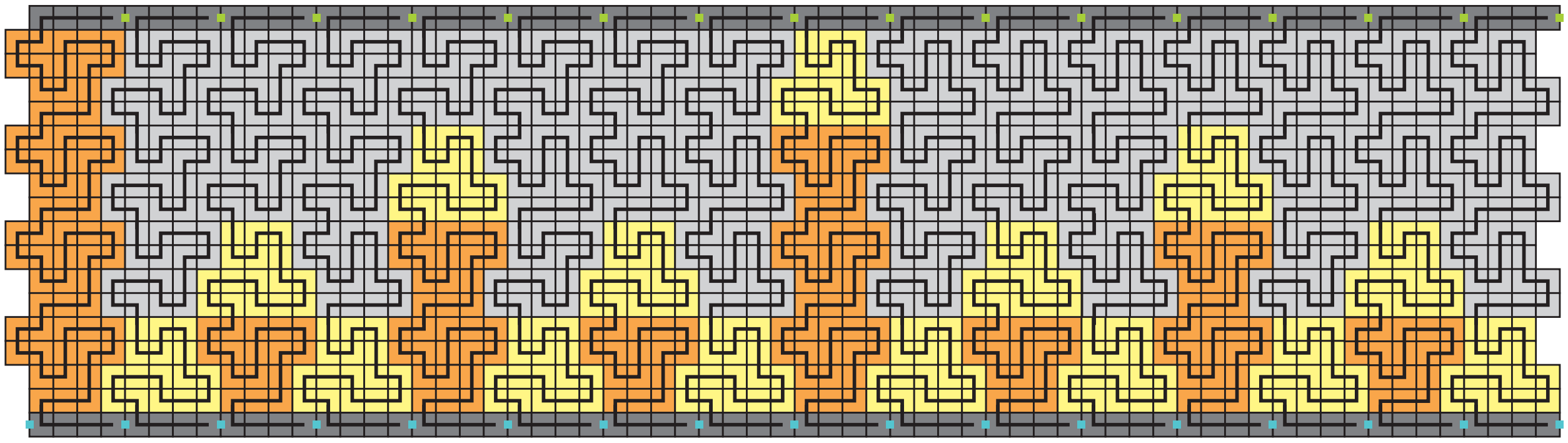

Figure 2: Fully assembled counter from the tile set given in Figure 1.

The glue types that connect blocks from one row to another ensure that any assembled column consists of red blocks from rows 1 to $r$ ($r$ at least 1 and at most $\log n$), followed by a yellow block in row $r+1$ (if $r < \log n$), followed by grey blocks (either type) in rows $r+2$ to row $\log n$ (if $r+1 < \log n$). This pattern has the property that for any $(\log n)$-bit string $b$, a column may assemble that encodes that string in the geometry of the dents on the left side of the column, and the right side of the column in turn encodes $b+1$. Additionally, a fully assembled column can also attach the two four-tile chains of Figure 1 to both the top $A$ glue and bottom $C$ glue of the column. For any two assembled columns, the strength $\tau = 1$ green and blue glues combined give a strength $\tau = 2$ affinity for any two assembled columns to attach. However, due to the rigid teeth-like geometry of the columns, only sequential columns can get close enough to realize the affinity and assemble under the two-handed assembly model. The unique assembly of the tile set of Figure 1 is shown in Figure 2.

In the example provided, we are specifically considering the special case of a counter that grows to a power of 2 length. More generally, it is possible to assemble only columns that encode values greater or equal to a given initial value, thereby allowing the assembly of a length-$n$ counter for general $n$. However, we leave these details for the extended fault tolerant version of the construction.

The counter in this section is not fuzzy fault tolerant. In particular, the supertile in Figure 3 is producible at temperature $\tau = 1$ (but not $\tau = 2$ because the two-handed model requires that at most 2 supertiles, both of which are stable at $\tau = 2$, combine in any step), stable at temperature $\tau = 2$, but cannot grow into the correct unique $\tau = 2$ assembled counter of Figure 2.

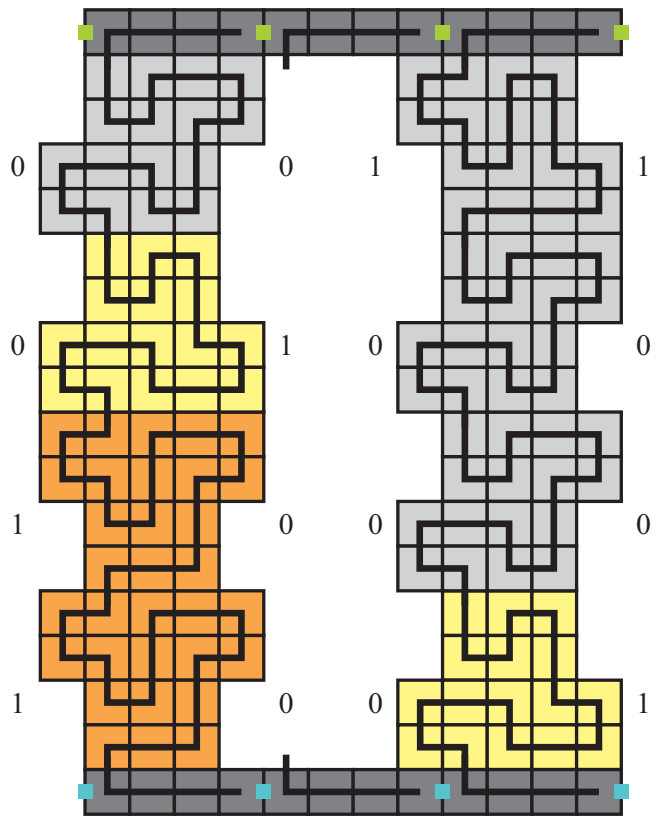

Figure 3: The basic temperature $\tau = 2$ counter in this section is not fuzzy fault tolerant. The above supertile is producible at temperature $\tau = 1$, stable at temperature $\tau = 2$, and cannot grow into the desired unique temperature $\tau = 2$ final assembly of Figure 2.

## 4 Fuzzy Temperature Fault-Tolerance

In this section we introduce the fuzzy temperature model of fault-tolerance in self-assembly. The fuzzy temperature assembly model permits rampant temperature $\tau = 1$ growth of supertiles under the two-handed assembly model. We are then interested in what producible temperature $\tau = 1$ assemblies become stable at temperature $\tau = 2$. If even a single temperature $\tau = 1$ assembly becomes stable at temperature $\tau = 2$ and is inconsistent with what can be built in a purely temperature $\tau = 2$ assembly model, the system is deemed *error prone*. On the other hand, if all temperature $\tau = 1$ assemblies that are stable at temperature $\tau = 2$ have a valid temperature $\tau = 2$ path of growth to a supertile that is producible under a pure temperature $\tau = 2$ model, then the system is deemed *fuzzy temperature fault-tolerant*. Put another way, even with arbitrary erroneous strength 1 attachments, a fuzzy temperature fault-tolerant system guarantees that such errors cannot stabilize at temperature 2 unless the stabilized supertile can itself grow into a *correct* temperature $\tau = 2$ assembly, which means such an assembly is not really an error.

Formally, for a given initial tile set $T$, we define fuzzy temperature fault-tolerance in terms of the following four sets of supertiles: (1) The *dependably produced* (DP) supertiles are those that can be assembled at temperature $\tau = 2$ under the two-handed assembly model. Formally, DP is the set of all producible supertiles for the two-handed assembly system $(T, 2)$; (2) The *dependably terminal* (DT) supertiles are all supertiles in DP that cannot grow any further at temperature $\tau = 2$. Formally, DT is the set of terminal, producible supertiles for the two-handed assembly system $(T, 2)$; (3) The *plausibly produced* (PP) supertiles are those that can be assembled at temperature $\tau = 1$. Formally, PP is the set of all producible supertiles for the two-handed assembly system $(T, 1)$; and (4) The *plausibly stable* (PS) supertiles are all supertiles in PP that are stable at temperature $\tau = 2$.

Intuitively, DT denotes a final collection of supertiles that can be expected to be built given enough time for assembly in a temperature 2 system. On the other hand, due to the occasional assembly of supertiles with only strength 1 attachments, elements in PP will (plausibly) be assembled. Elements of PP that are not stable at temperature $\tau = 2$ intuitively will eventually break apart and are not of concern. However, these assemblies may grow to a point in which they become stable at temperature $\tau = 2$, in which case they will not break apart. Such assemblies constitute the set PS. The goal is to design a system such that for each element $\alpha$ of PS, every terminal $\beta$ into which $\alpha$ can grow at temperature $\tau = 2$ is an element of DT (written $PS \Rightarrow DT$), and that DT is the set of desired shapes to be assembled. Put another way, we want to avoid the design of an error prone system in which stable assemblies that are inconsistent with the desired final assembly are built by erroneous $\tau = 1$ strength attachments.

More precisely, the fuzzy temperature fault-tolerance design problem is as follows:

**fuzzy temperature fault-tolerance design problem:** Given a target shape $\Upsilon$, the goal is to design a tile set such that: (1) $PS \Rightarrow DT$ (fuzzy temperature fault-tolerance constraint); and (2) all supertiles in DT have shape $\Upsilon$. (Desired goal shape is the unique output of the assembly.)

For the remainder of this paper, we attempt to solve the fuzzy temperature fault-tolerance problem for the benchmark example of an $n \times n$ square. As a metric, we are interested in minimizing the number of distinct tile types required to assemble a square while adhering to the fuzzy temperature fault-tolerance constraint; the problem is trivialized if one allows $n^2$ different tile types to hard-code each position in the square (or even using $O(n)$ tile types to use the non-cooperative "comb" structure from [34]). We show that a sleek $O(\log n)$ tile complexity is achievable, which is very close to the $O\left(\frac{\log n}{\log\log n}\right)$ bound that can be achieved with no fault-tolerance constraint (in the seeded, single-tile addition model).

# 5 Overview of Fault-Tolerant Square Construction

This section gives a high-level description of the main construction of this paper, a square that assembles under the fuzzy temperature fault tolerance model. A more detailed description of the construction can be found in Section B.

## 5.1 Square

As is common in many self-assembly constructions for square-building, most of the work is in constructing counters that calculate the dimensions of the square. Figure 4 shows a high-level diagram of how to compose these counters. The horizontal counter and the vertical counters are constructed in conceptually the same way, with minor differences in the actual implementation. Most of the effort of our main construction is in encoding the number $n$ into the tiles that grow a counter, so that it can control the length to which the counter grows, in a fault-tolerant way.

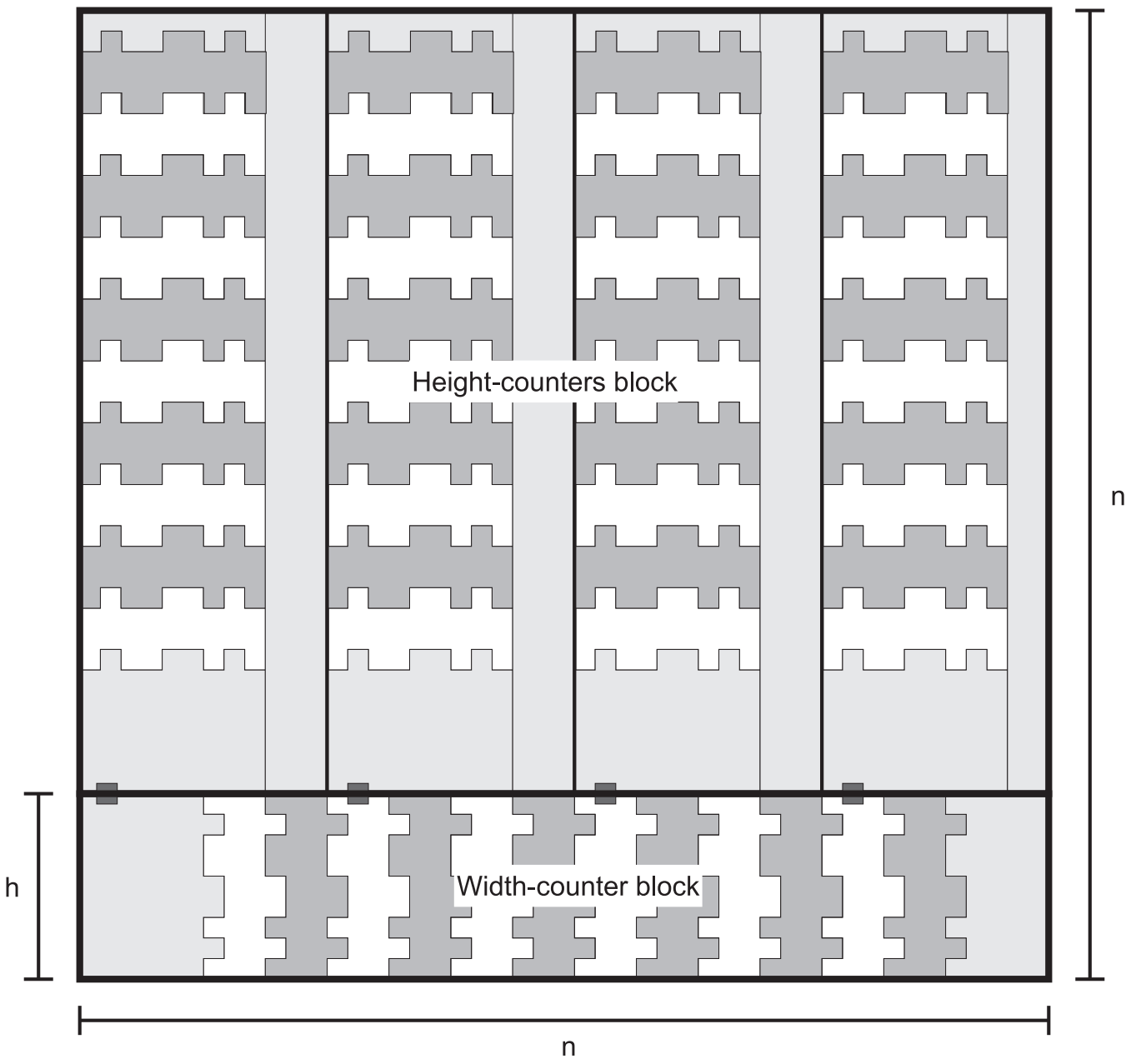

Figure 4: A simplified diagram of the components of a full $n \times n$ square. Components are not represented to scale.

## 5.2 Counter

For simplicity we describe only the horizontal counter. The vertical counters are constructed similarly, with the exception that they are slightly simpler because of the need for the horizontal counter to correctly space out its bonds designed to connect the horizontal counter to the various vertical counters.

Define $k \equiv \lfloor \log n \rfloor + 2$ to be 1 plus the number of bits in $n$. As in Section 3, the counter consists of $\approx n$ columns (actually $n$ divided by the width in tiles of a column, which is a constant, but for simplicity of discussion we will assume that there are $n$ columns), each representing an integer between $2^k - n$ and $2^k - 1$. Note that we refer to columns as "counter-values." Each counter-value is connected to the next by two strength-1 *inter-counter-value glues*, and correct inter-counter-value binding is enforced using bumps and dents as in Section 3.

## 5.3 Counter-Value

As in Section 3, counter-values form randomly from $\approx \log n$ "bit gadgets", each of constant size, with each bit selected at random. Figure 6 shows the bit gadgets, and Figure 7 shows some of them attaching to form a few counter-values of a counter. Beyond the need for fuzzy temperature fault-tolerance, these bit gadgets must meet additional requirements. We first describe how to meet these requirements, and then describe how to achieve fault-tolerance.

### 5.3.1 Glue Design for Additional Requirements of Counter-Values

The logical requirements that counter-values must meet are:

(a) The right side of a counter-value must represent $i + 1$ if the left side represents $i$. This was already needed in Section 3.

(b) Each counter-value must be guaranteed to form an integer in the range $\left[2^k - n, 2^k - 1\right]$, so that the counter has exactly $n$ counter-values.

(c) Only a subset of appropriately spaced counter-values should have glues on the north to allow the vertical counters to bind, since the horizontal width of each vertical counters is $\Theta(\log n)$, whereas the horizontal width of each counter-value in the horizontal counter is $O(1)$. This is done by choosing a power of two $2^m$ (for $m$ just large enough that $2^m >$ width of a vertical counter), and placing the glues to the north every $2^m$ counter-values.

The fault-tolerance is achieved entirely through the geometric design of the bit gadgets, and the choice of binding paths within them. The requirements (a), (b), and (c) are achieved through careful selection of the north-south glues that connect bit gadgets to each other. For the sake of meeting these three requirements, we can therefore logically view each bit gadget as a single tile, with double-strength glues on the north and south. The values of these glues will then be carried through to every actual tile that makes up a bit gadget, and combined with the glues that hard-code the relative position of each tile in the bit gadget, allow us to conceptually separate the problem of fault tolerance from that of meeting the three requirements discussed above. Finally, we can conceptually separate these three problems from each other, designing tiles to meet those requirements separately, and combine them in a cross-product construction. Figure 5 shows the three tile sets that meet the requirements (a), (b), and (c). (A brief discussion follows, but see Section B.2 for an in-depth description.)

In each case, we take care to ensure that the requirement is met no matter in which order the tiles aggregate. Nonetheless, it is easiest to describe their operation as though the northmost tile

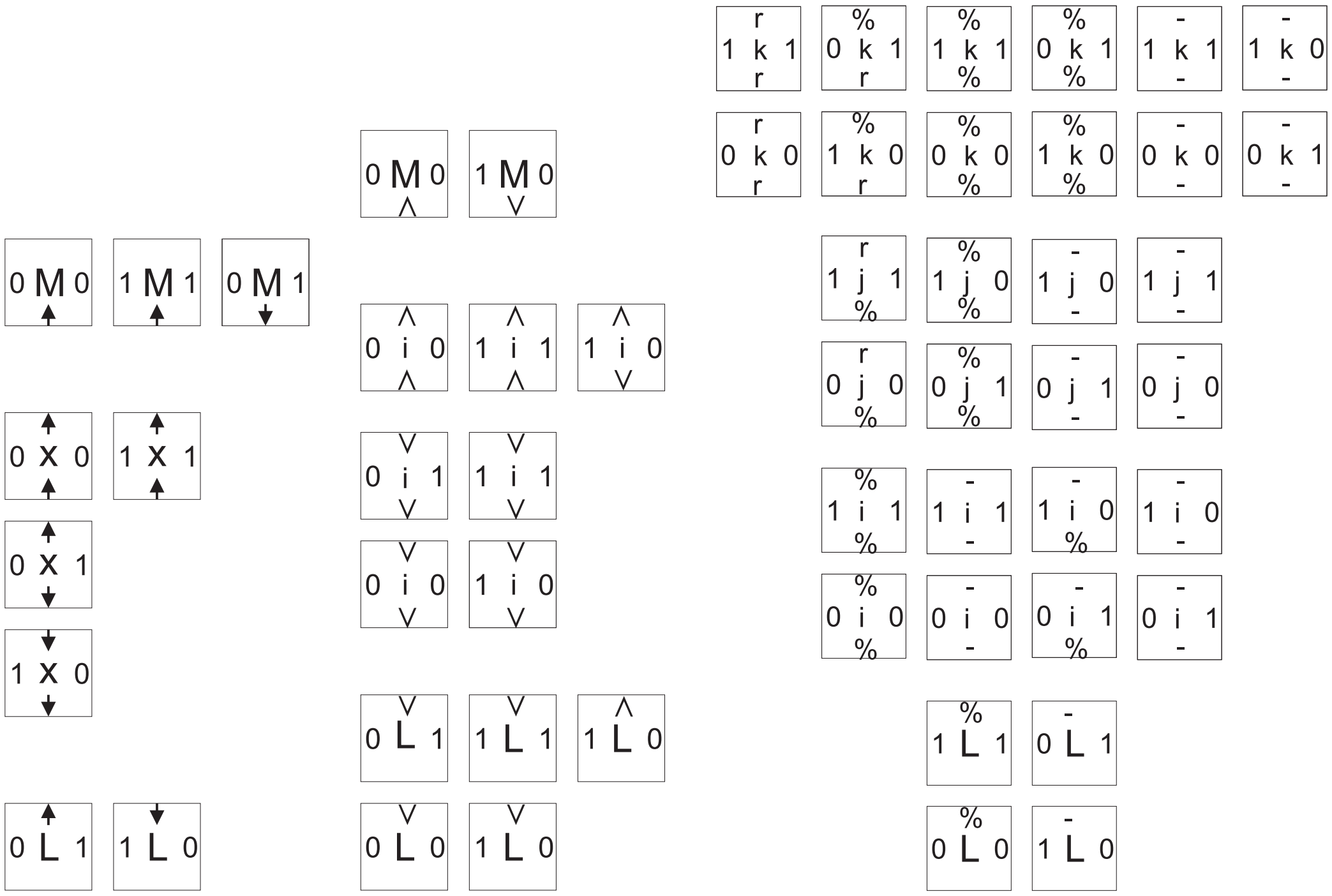

(a) Increment a binary number

(b) Enforces that the left number is greater than the right

(c) Detects if the binary number formed 1) is equal to a given number, or 2) matches a given number of the low-order bits of that number

Figure 5: Templates for tile sets that perform subsets of the functionality of the hairpin gadgets (described in more detail in Section 5.3.2). Though not shown, each tile has strength-2 glues on the north and south, implemented in the actual tile set as a pair of single-strength bonds for fault-tolerance purposes. The east and west "bit values" in Figure 5a are represented in the actual tile set by the geometric shape of the $16 \times 16$ tile bit gadget that each individual tile in this figure represents, geometrically enforcing agreement on the bits of adjacent bit gadgets.

is first present, and the counter-value assembles north-to-south; i.e., most significant bit to least significant.

Figure 5a shows the tiles that implement incrementing to ensure that the east bits represent $i+1$ if the west bits represent $i$. If the position of the least significant 0 in $i$ is $p$, then all bits at positions above $p$ are equal, all bits at positions below $p$ are 1 for $i$ and 0 for $i+1$, and at position $p$ the bit is 0 for $i$ and 1 for $i+1$. Therefore the tiles nondeterministically guess a position $p$ at which to make this transition, and enforce that all tiles above $p$ have equal bits and all tiles at or below $p$ obey the stated requirement.

Figure 5b shows the tiles that implement range-checking to ensure that the number $i$ that is constructed is greater than $m = 2^k - n$. (Since precisely $k$ bits are assembled, $i < 2^k$.) Imagine comparing $i$ to $m$ starting at the most significant bit. We must enforce that there is at least one bit difference, and that in the position of most significance where there is a difference that the bit from $i$ is 1 and the bit from $m$ is 0. As before, the tiles nondeterministically guess at which position the first disagreement will occur. Below the first disagreement, the bits of $i$ are selected nondeterministically. We chose the value of $k$ so that we know $n$'s most significant bit is 0; this helps to ensure, if tiles grow from south to north and have not yet enforced $i > m$, then the most significant bit of $i$ may be chosen equal to 1 to enforce this.

Figure 5c shows the tiles that ensure that two single-strength glues designed to be an anchor point for vertical counters are placed on the top of a counter-value in the horizontal counter if and only if the counter-value is at an appropriate position to space the vertical counters out evenly. This is accomplished by first determining the number, $r$, which will be represented by the rightmost counter-value for which this will be the case. Then, whenever the number $i$ represented by a counter-value shares the same least significant $m$ bits with $r$, the northern glues are present to anchor a vertical counter. Additionally, in the special case where all bits of $i$ match those of $r$, a pair of northern glues unique to that position are present, to ensure that the special case, rightmost vertical counter with the necessary padding to fill out the width to exactly $n$, can attach.

### 5.3.2 Geometric Design for Fault-Tolerance

On the assumption that the three requirements in the previous section can be met for each counter-value that forms, we now describe how to use geometry and "synchronization primitives" involving careful placement of glues to ensure that even at temperature 1, unwanted structures cannot grow that will be stable at temperature 2. Recall that at temperature 2, the counter-values of the counter of Section 3 enforce that binding between adjacent counter-values cannot occur until both counter-values are fully assembled; this occurs because the path (consisting of all strength-2 glues) from one single-strength inter-counter-value glue to another goes through every bump of the counter-value. Hence, to have both glues present, the entire counter-value must also be present.

Our construction enforces that no structure producible even at temperature 1 can stably attach to the east of counter-value $i$ unless it contains enough of the bumps of its westmost counter-value to enforce that binding requires that counter-value to represent $i + 1$. This is enforced by the following constraint: every path (including strength-1 glues) connecting the two inter-counter-value glues of counter-value $i$ that intersects any counter-value $j > i$, also passes through every bump of the counter-value $i + 1$. Therefore, enough of the leftmost counter-value of this structure is guaranteed to be present to ensure that it can only bind to the right of counter-value $i$ if its leftmost counter-value represents $i + 1$.

To enforce that a path from some part of counter-value $i$ to some part of counter-value $i+2$ must traverse the entire height of counter-value $i + 1$, we must enforce that a path traverses southward through the bumps of counter-value $i + 1$, and then traverses northward again before moving on to counter-value $i + 2$. But since the path cannot "short-circuit" there must be no glues between the southward and northward paths except at the bottom of the counter-value. The bumps and dents on the east side of the southward path must be faithfully represented on the east side of the northward path.

Even though the bits can grow in any order, it is easiest to imagine growing the bits of the southward path, then turning around and guessing those same bits while growing the northward path. Each bit along a single path is represented by what we will call a *hairpin gadget*; one southward and one northward hairpin gadget (though unconnected to each other) form a single bit gadget. To ensure that improper guesses do not result in junk assemblies that cannot grow any further, we use a similar motif to the "single-strength glues at opposite ends" used in Section 3, within the hairpin gadgets themselves. That is, hairpin gadgets can only bind stably to the north of other hairpin gadgets when fully formed, which prevents a hairpin gadget that does not match its complementary hairpin gadget from locking in. Figure 6 shows the individual hairpin gadgets along with gadgets designed to secure them to each other. Figure 7 shows part of a counter formed from these gadgets. The white hairpin gadgets are "southward growing" (again, if we imagine tracing a path from counter-value $i$ to counter-value $i + 1$, bearing in mind that the two-handed assembly can grow in other orders), and the gray hairpin gadgets are "northward growing".

Intuitively, the only connection between a white "left-half" of a counter-value and the gray "right-half" of that counter-value is through the southern row. Northward growth from this row is kept consistent by ensuring that no hairpin gadget can stabilize to the hairpin gadget beneath it until the red double-bond is present. Since every path from this red double-bond to a blue single-bond on the south of the same hairpin gadget goes through the bumps of that gadget, the gadget cannot stabilize unless it is consistent with what has already grown to the left or right of it (and if nothing has already, then *it* determines what must be consistent with it).

Conversely, southward growth, which can lock a hairpin gadget to the hairpin gadget to its north *without* necessarily agreeing with the hairpin to its left or right, nevertheless cannot stabilize at temperature 2 without growing enough of those bumps to enforce agreement. This is because the bottom row must be present to connect a white counter-value half to its gray half, and both must be present to connect that counter-value to the previous (left) counter-value.

Finally, the bumps and dents at the eastmost and westmost edges leave some space that must be filled in if we wish to create a true square, and to create an exactly $n \times n$ square for any $n$ there is additional 'padding' necessary. In order to obtain smooth edges for the counter, the first and last counter-value assemblies are created from hard-coded sets of tiles which have no bumps and dents on their left and right sides, respectively. Finally, additional hard-coded rows of the necessary length for the full width to be exactly $n$ attach to the right side of the last counter value. (An example of such padding is shown in Figure 8.) In a similar way, the spacing between vertical counters is filled in.

## 6 Conclusion

Adleman, Cheng, Goel, and Huang [3] show that for each $n$ there is a (seeded, single-tile addition, non-fault-tolerant) tile assembly system that uniquely assembles an $n \times n$ square using $O(\log n / \log \log n)$ unique tile types, a bound that was shown asymptotically tight by Rothemund and Winfree [34]. Since our construction uses $\Theta(\log n)$ tile types, an obvious open question is whether there is a fuzzy temperature fault-tolerant tile assembly system that uses the asymptotically optimal $O(\log n / \log \log n)$ to uniquely assemble an $n \times n$ square. Previous papers [2–4] have focused on running time for self-assembled shapes. This is a particularly difficult problem for two-handed assembly. The papers attacking the case of the two-handed model [2,4] expend much effort to derive the expected assembly time for the much simpler problem of assembling a 1-dimensional $1 \times n$ line from $n$ unique tile types that each encode a different position in the line. It is an open problem, first stated in [2], to prove upper or lower bounds for the optimal time to assemble a square under the two-handed model. It is also an open problem to derive the expected time to completion for our more complicated construction of a fuzzy temperature fault-tolerant square.

Our construction is "floppy": many adjacent tiles in the final square are not connected by glues. One would expect that more strongly connected squares are more physically resilient, and they may also help to enforce the steric protection utilized in our construction, so this floppiness may be a disadvantage. Given the goal of preventing all erroneous temperature-1 growth from stabilizing, it seems unlikely that a *full square* – a square in which every neighboring pair of tiles interact with positive strength – could be constructed using a fuzzy temperature fault-tolerant system. But it is conceivable that more elaborate use of synchronization could allow extra "support substructures" to be used to make our construction "more fully connected", while preserving the fuzzy temperature fault-tolerance.

Additionally, the two-handed aTAM, while more realistic than the seeded single-tile attachment aTAM in the sense that it allows for nucleation without a seed, is perhaps less realistic in another sense. The DNA tiles that the aTAM was originally conceived to model, while ostensibly two-dimensional, are not necessarily confined to the plane. In particular, the steric protection that we employ requires the tiles in the $x$-$y$ plane to stay at position $z = 0$. If two mismatching gadgets collide, but one of them "slides" over the other by moving its bumps out of the plane into $z > 0$, then this could allow the cooperative strength-1 bonds to connect even between mismatching gadgets. We note that other theoretical papers require the assumption of planar steric hindrance as well [1, 5, 14, 16], so this potential shortcoming is not unique to our technique.

However, floppiness is not an unbreakable law of physics; it is an artifact of one particular experimental method of using DNA to create self-assembling tiles. It is not necessarily infeasible to construct tiles by another method that stay in the plane, or thicken them along the $z$-axis so that some floppiness is tolerable while still enforcing blocking due to steric protection. There are macro-scale techniques for tile self-assembly that are more sturdy and likely to stay in the plane [9,32], as well as nanoscale techniques for creating rigid DNA structures [18,25]. Nonetheless, since a promising current technology for constructing self-assembling molecular tiles is the DNA double-crossover implementation, floppiness is a potential problem, for our construction as well as the other theoretical constructions that utilize planar blocking [1, 5, 14, 16]. It remains an open theoretical problem to design a construction of a fuzzy-temperature fault-tolerant square from $O(\log n)$ tile types that is robust to "3-D floppiness", and an open experimental problem to design physical molecular tiles that are inflexible enough to allow the use of programmed steric protection as a reliable design tool. Another open experimental problem in two-handed tile assembly is to determine, for a given tile implementation, what is the largest size of supertiles that will reliably combine. While it is clear that single tiles experience enough motion in solution to move into positions necessary to combine to growing assemblies, and most likely that supertiles consisting of small numbers of tiles will also do so, there may be an upper bound on the size of supertiles that reliably attach. Note that this potential limitation would also apply to other two-handed constructions employing arbitrarily large supertiles [2, 4, 6, 14].

**Acknowledgments.** We thank anonymous referees for suggested improvements to this paper.

# A Two-Handed Abstract Tile Assembly Model

In this section we formally define a variant of Erik Winfree's abstract Tile Assembly Model [43,44] modified to model unseeded growth, known as the *two-handed* aTAM, which has been studied previously under various names [2,4,6,14,27,45]. In the two-handed aTAM, any two assemblies can attach to each other, rather than enforcing that tiles can only accrete one at a time to an existing seed assembly. In this section we define the model to allow for infinite assemblies and systems that produce more than one terminal assembly, even though our main construction does not have these properties. We also allow for systems that start with a finite number of tile types, even though the description of our main construction is in terms of an infinite supply of tiles.

$\lg(x)$ and $\log(x)$ each denote the base-2 logarithm of $x$, and $\ln(x)$ denotes the base-$e$ logarithm of $x$. We work in the 2-dimensional discrete space $\mathbb{Z}^2$. Define the set $U_2 = \{(0,1),(1,0),(0,-1),(-1,0)\}$ to be the set of all *unit vectors*, i.e., vectors of length 1 in $\mathbb{Z}^2$. All *graphs* in this paper are undirected. A *grid graph* is a graph $G = (V,E)$ in which $V \subseteq \mathbb{Z}^2$ and every edge $\{\vec{a},\vec{b}\} \in E$ has the property that $\vec{a} - \vec{b} \in U_2$.

Intuitively, a tile type $t$ is a unit square that can be translated, but not rotated, having a well-defined "side $\vec{u}$" for each $\vec{u} \in U_2$. Each side $\vec{u}$ of $t$ has a "glue" with "label" $\text{label}_t(\vec{u})$ – a string over some fixed alphabet – and "strength" $\text{str}_t(\vec{u})$ – a nonnegative integer – specified by its type $t$. Two tiles $t$ and $t'$ that are placed at the points $\vec{a}$ and $\vec{a}+\vec{u}$ respectively, *bind* with *strength* $\text{str}_t(\vec{u})$ if and only if $(\text{label}_t(\vec{u}), \text{str}_t(\vec{u})) = (\text{label}_{t'}(-\vec{u}), \text{str}_{t'}(-\vec{u}))$.

In the subsequent definitions, given two partial functions $f, g$, we write $f(x) = g(x)$ if $f$ and $g$ are both defined and equal on $x$, or if $f$ and $g$ are both undefined on $x$.

Throughout this section, fix a finite set $T$ of tile types. An *assembly* is a partial function $\alpha : \mathbb{Z}^2 \dashrightarrow T$ defined on at least one input, with points $\vec{x} \in \mathbb{Z}^2$ at which $\alpha(\vec{x})$ is undefined interpreted to be empty space, so that $\text{dom}\,\alpha$ is the set of points with tiles. We write $|\alpha|$ to denote $|\text{dom}\,\alpha|$, and we say $\alpha$ is *finite* if $|\alpha|$ is finite. For assemblies $\alpha$ and $\alpha'$, we say that $\alpha$ is a *subassembly* of $\alpha'$, and write $\alpha \sqsubseteq \alpha'$, if $\text{dom}\,\alpha \subseteq \text{dom}\,\alpha'$ and $\alpha(\vec{x}) = \alpha'(\vec{x})$ for all $x \in \text{dom}\,\alpha$. Two assemblies $\alpha$ and $\beta$ are *disjoint* if $\text{dom}\,\alpha \cap \text{dom}\,\beta = \varnothing$. For two assemblies $\alpha$ and $\beta$, define the *union* $\alpha \cup \beta$ to be the assembly defined for all $\vec{x} \in \mathbb{Z}^2$ by $(\alpha \cup \beta)(\vec{x}) = \alpha(\vec{x})$ if $\alpha(\vec{x})$ is defined, and $(\alpha \cup \beta)(\vec{x}) = \beta(\vec{x})$ otherwise. Say that this union is *disjoint* if $\alpha$ and $\beta$ are disjoint.

The *binding graph of* an assembly $\alpha$ is the grid graph $G_\alpha = (V,E)$, where $V = \text{dom}\,\alpha$, and $\{\vec{m},\vec{n}\} \in E$ if and only if (1) $\vec{m} - \vec{n} \in U_2$, (2) $\text{label}_{\alpha(\vec{m})}(\vec{n} - \vec{m}) = \text{label}_{\alpha(\vec{n})}(\vec{m} - \vec{n})$, and (3) $\text{str}_{\alpha(\vec{m})}(\vec{n} - \vec{m}) > 0$. Given $\tau \in \mathbb{N}$, an assembly is $\tau$*-stable* (or simply *stable* if $\tau$ is understood from context), if it cannot be broken up into smaller assemblies without breaking bonds of total strength at least $\tau$; i.e., if every cut of $G_\alpha$ has weight at least $\tau$, where the weight of an edge is the strength of the glue it represents. In contrast to the model of Wang tiling, the nonnegativity of the strength function implies that glue mismatches between adjacent tiles do not prevent a tile from binding to an assembly, so long as sufficient binding strength is received from the (other) sides of the tile at which the glues match.

The two-handed aTAM [6,14] allows for two assemblies, both possibly consisting of more than one tile, to attach to each other, in contrast to the seeded aTAM [34,44] in which one of the attaching objects is assumed to be a single tile, and the other is the assembly containing the unique "seed tile". Since we must allow that the assemblies might require translation before they can bind, we define a *supertile* to be the set of all translations of a $\tau$-stable assembly, and speak of the attachment of supertiles to each other, modeling that the assemblies attach, if possible, after appropriate translation.

Formally, for assemblies $\alpha, \beta : \mathbb{Z}^2 \dashrightarrow T$ and $\vec{u} \in \mathbb{Z}^2$, we write $\alpha + \vec{u}$ to denote the assembly defined for all $\vec{x} \in \mathbb{Z}^2$ by $(\alpha+\vec{u})(\vec{x}) = \alpha(\vec{x}-\vec{u})$, and write $\alpha \simeq \beta$ if there exists $\vec{u}$ such that $\alpha+\vec{u} = \beta$; i.e., if $\alpha$ is a translation of $\beta$. Define the *supertile* of $\alpha$ to be the set $\tilde{\alpha} = \{ \beta \mid \alpha \simeq \beta \}$. A supertile $\tilde{\alpha}$ is *$\tau$-stable* (or simply *stable*) if all of the assemblies it contains are $\tau$-stable; equivalently, $\tilde{\alpha}$ is stable if it contains a stable assembly, since translation preserves the property of stability. Note also that the notation $|\tilde{\alpha}| \equiv |\alpha|$ is well-defined, since translation preserves cardinality (and note in particular that even though we define $\tilde{\alpha}$ as a set, $|\tilde{\alpha}|$ does not denote the cardinality of this set, which is always $\aleph_0$).

For two supertiles $\tilde{\alpha}$ and $\tilde{\beta}$, and temperature $\tau \in \mathbb{N}$, define the *combination* set $C^{\tau}_{\tilde{\alpha},\tilde{\beta}}$ to be the set of all supertiles $\tilde{\gamma}$ such that there exist $\alpha \in \tilde{\alpha}$ and $\beta \in \tilde{\beta}$ such that (1) $\alpha$ and $\beta$ are disjoint, (2) $\gamma \equiv \alpha \cup \beta$ is $\tau$-stable, and (3) $\gamma \in \tilde{\gamma}$. That is, $C^{\tau}_{\tilde{\alpha},\tilde{\beta}}$ is the set of all $\tau$-stable supertiles that can be obtained by attaching $\tilde{\alpha}$ to $\tilde{\beta}$ stably, with $|C^{\tau}_{\tilde{\alpha},\tilde{\beta}}| > 1$ if there is more than one position at which $\beta$ could attach stably to $\alpha$.

It is common with seeded assembly to stipulate an infinite number of copies of each tile, but our definition allows for a finite number of tiles as well. Our definition also allows for the growth of infinite assemblies and finite assemblies to be captured by a single definition, similar to the definitions of [24] for seeded assembly.

Given a set of tiles $T$, define a *state* $S$ of $T$ to be a multiset of supertiles, or equivalently, $S$ is a function mapping supertiles of $T$ to $\mathbb{N} \cup \{\infty\}$, indicating the multiplicity of each supertile in the state. We therefore write $\tilde{\alpha} \in S$ if and only if $S(\tilde{\alpha}) > 0$.

A *(two-handed) tile assembly system* (*TAS*) is an ordered triple $\mathcal{T} = (T, S, \tau)$, where $T$ is a finite set of tile types, $S$ is the *initial state*, and $\tau \in \mathbb{N}$ is the temperature. Subsequently we assume that $\tau = 2$, unless explicitly stated otherwise. If not stated otherwise, assume that the initial state $S$ is defined $S(\tilde{\alpha}) = \infty$ for all supertiles $\tilde{\alpha}$ such that $|\tilde{\alpha}| = 1$, and $S(\tilde{\beta}) = 0$ for all other supertiles $\tilde{\beta}$. That is, $S$ is the state consisting of a countably infinite number of copies of each individual tile type from $T$, and no other supertiles. In such a case we write $\mathcal{T} = (T, \tau)$ to indicate that $\mathcal{T}$ uses the default initial state.

Given a TAS $\mathcal{T} = (T, S, \tau)$, define an *assembly sequence* of $\mathcal{T}$ to be a sequence of states $\vec{S} = (S_i \mid 0 \leq i < k)$ (where $k = \infty$ if $\vec{S}$ is an infinite assembly sequence), and $S_{i+1}$ is constrained based on $S_i$ in the following way: There exist supertiles $\tilde{\alpha}, \tilde{\beta}, \tilde{\gamma}$ such that (1) $\tilde{\gamma} \in C^{\tau}_{\tilde{\alpha},\tilde{\beta}}$, (2) $S_{i+1}(\tilde{\gamma}) = S_i(\tilde{\gamma}) + 1$,[2] (3) if $\tilde{\alpha} \neq \tilde{\beta}$, then $S_{i+1}(\tilde{\alpha}) = S_i(\tilde{\alpha}) - 1$, $S_{i+1}(\tilde{\beta}) = S_i(\tilde{\beta}) - 1$, otherwise if $\tilde{\alpha} = \tilde{\beta}$, then $S_{i+1}(\tilde{\alpha}) = S_i(\tilde{\alpha}) - 2$, and (4) $S_{i+1}(\tilde{\omega}) = S_i(\tilde{\omega})$ for all $\tilde{\omega} \notin \{\tilde{\alpha}, \tilde{\beta}, \tilde{\gamma}\}$. That is, $S_{i+1}$ is obtained from $S_i$ by picking two supertiles from $S_i$ that can attach to each other, and attaching them, thereby decreasing the count of the two reactant supertiles and increasing the count of the product supertile. If $S_0 = S$, we say that $\vec{S}$ is *nascent*.

Given an assembly sequence $\vec{S} = (S_i \mid 0 \leq i < k)$ of $\mathcal{T} = (T, S, \tau)$ and a supertile $\tilde{\gamma} \in S_i$ for some $i$, define the *predecessors* of $\tilde{\gamma}$ in $\vec{S}$ to be the multiset $\text{pred}_{\vec{S}}(\tilde{\gamma}) = \{\tilde{\alpha}, \tilde{\beta}\}$ if $\tilde{\alpha}, \tilde{\beta} \in S_{i-1}$ and $\tilde{\alpha}$ and $\tilde{\beta}$ attached to create $\tilde{\gamma}$ at step $i$ of the assembly sequence, and define $\text{pred}_{\vec{S}}(\tilde{\gamma}) = \{\tilde{\gamma}\}$ otherwise. Define the *successor* of $\tilde{\gamma}$ in $\vec{S}$ to be $\text{succ}_{\vec{S}}(\tilde{\gamma}) = \tilde{\alpha}$ if $\tilde{\gamma}$ is a predecessor of $\tilde{\alpha}$ in $\vec{S}$, and define $\text{succ}_{\vec{S}}(\tilde{\gamma}) = \tilde{\gamma}$ otherwise. A sequence of supertiles $\vec{\tilde{\alpha}} = (\tilde{\alpha}_i \mid 0 \leq i < k)$ is a *supertile assembly sequence* of $\mathcal{T}$ if there is an assembly sequence $\vec{S} = (S_i \mid 0 \leq i < k)$ of $\mathcal{T}$ such that, for all $1 \leq i < k$, $\text{succ}_{\vec{S}}(\tilde{\alpha}_{i-1}) = \tilde{\alpha}_i$, and $\vec{\tilde{\alpha}}$ is *nascent* if $\vec{S}$ is nascent.

[2]with the convention that $\infty = \infty + 1 = \infty - 1$

The *result* of a supertile assembly sequence $\vec{\tilde{\alpha}}$ is the unique supertile $\text{res}(\vec{\tilde{\alpha}})$ such that there exist an assembly $\alpha \in \text{res}(\vec{\tilde{\alpha}})$ and, for each $0 \leq i < k$, assemblies $\alpha_i \in \tilde{\alpha}_i$ such that $\text{dom}\,\alpha = \bigcup_{0\leq i<k} \text{dom}\,\alpha_i$ and, for each $0 \leq i < k$, $\alpha_i \sqsubseteq \alpha$. For all supertiles $\tilde{\alpha}, \tilde{\beta}$, we write $\tilde{\alpha} \to_{\mathcal{T}} \tilde{\beta}$ (or $\tilde{\alpha} \to \tilde{\beta}$ when $\mathcal{T}$ is clear from context) to denote that there is a supertile assembly sequence $\vec{\tilde{\alpha}} = (\tilde{\alpha}_i \mid 0 \leq i < k)$ such that $\tilde{\alpha}_0 = \tilde{\alpha}$ and $\text{res}(\vec{\tilde{\alpha}}) = \tilde{\beta}$. It can be shown using the techniques of [33] for seeded systems that for all two-handed tile assembly systems $\mathcal{T}$ supplying an infinite number of each tile type, $\to_{\mathcal{T}}$ is a transitive, reflexive relation on supertiles of $\mathcal{T}$.

A supertile $\tilde{\alpha}$ is *producible*, and we write $\tilde{\alpha} \in \mathcal{A}[\mathcal{T}]$, if it is the result of a nascent supertile assembly sequence. A supertile $\tilde{\alpha}$ is *terminal* if, for all producible supertiles $\tilde{\beta}$, $C^{\tau}_{\tilde{\alpha},\tilde{\beta}} = \varnothing$.[3] Define $\mathcal{A}_{\Box}[\mathcal{T}] \subseteq \mathcal{A}[\mathcal{T}]$ to be the set of terminal and producible supertiles of $\mathcal{T}$. $\mathcal{T}$ is *directed* (a.k.a., *deterministic*, *confluent*) if $|\mathcal{A}_{\Box}[\mathcal{T}]| = 1$.

Let $X \subseteq \mathbb{Z}^2$ be a shape. We say $X$ *strictly self-assembles* in $\mathcal{T}$ if, for all $\tilde{\alpha} \in \mathcal{A}_{\Box}[\mathcal{T}]$, there exists $\alpha \in \tilde{\alpha}$ such that $\text{dom}\,\alpha = X$; i.e., $\mathcal{T}$ uniquely assembles into the shape $X$.

# B Fault-Tolerant Assembly of a Square with $O(\log n)$ Tile Types

Figure 4 shows a high-level depiction of the main logical components, which create an $n \times n$ square. The square consists of two logical components: a bottom portion, which we will call the "width-counter block," and a top portion called the "height-counters block." The width-counter block is essentially a counter that calculates the width of the square. This counter is composed of a series of fixed width (16 tiles wide) columns, each representing a binary value $x$ on its left side and the value $x + 1$ in its right side. The height-counter blocks are composed of a series of vertically oriented counters (similar to the horizontal counter in the width-counter block) that count to a lesser value so that their height (with the necessary padding) is $n$ minus the height of the width-counter block. These counters are designed so that they attach to the north side of the width-counter block at specified locations.

Below are quantities that (collectively) define the dimensions and attachment locations of various sub-assemblies, the combinations of which form the completed square.

- $n$: the value of the height and width of the square to be assembled
- 16: the width and height of a "bit gadget," which is a sub-assembly representing a single bit in a value of a counter, this consists of the combination of a white hairpin gadget and its complementary grey hairpin gadget (see Figure 6). Note that the bit gadgets for the most significant bits are 16 tiles wide but only 13 tiles tall in order to compensate for the 2 extra rows of tiles on the top and 1 on the bottom which are used to connect the counter values.
- $c^{\to} = \lfloor n/16 \rfloor$, the number of counter values, or binary values to be counted by the width-counter, where a counter value is represented by the vertical combination of bit gadgets. $O(\lg(n))$ tile types are required to form these counter values.

[3]Note that a supertile $\tilde{\alpha}$ could be non-terminal in the sense that there is a producible supertile $\tilde{\beta}$ such that $C^{\tau}_{\tilde{\alpha},\tilde{\beta}} \neq \varnothing$, yet it may not be possible to produce $\tilde{\alpha}$ and $\tilde{\beta}$ simultaneously if some tile types are given finite initial counts, implying that $\tilde{\alpha}$ cannot be "grown" despite being non-terminal. If the count of each tile type in the initial state is $\infty$, then all producible supertiles are producible from any state, and the concept of terminal becomes synonymous with "not able to grow", since it would always be possible to use the abundant supply of tiles to assemble $\tilde{\beta}$ alongside $\tilde{\alpha}$ and then attach them.

- $p^{\rightarrow} = n - 16c^{\rightarrow}$, the extra width, or padding, added to the rightmost counter value of the width-counter to ensure that the width of the square is exactly $n$ (the range is 0-15, thus requiring a constant sized set of tile types which form rows of length $p^{\rightarrow}$ that attach to the right edge of the rightmost counter value of the width-counter. See Figure 8 for an example.)
- $k^{\rightarrow} = 2^{\lceil \lg(c^{\rightarrow}) \rceil} - c^{\rightarrow}$, the starting counter value for the width-counter, which will count from $k^{\rightarrow}$ to the next power of two, $2^{\lceil \lg(c^{\rightarrow}) \rceil}$. The counter value $k^{\rightarrow}$ is actually formed by a hard-coded set of tiles (requiring $O(\lg(n))$ tile types), and the first counter value which actually assembles from bit gadgets is $k^{\rightarrow} + 1$.
- $k^{\rightarrow}_{Max} = 2^{\lceil \lg(c^{\rightarrow}) \rceil} - 1$, the maximum counter value reached by the width-counter (consisting of all 1's in binary). Similar to the counter value for $k^{\rightarrow}$, this counter value is also represented by a hard-coded assembly requiring $O(\lg(n))$ tile types and which also has glues on the right side that allow the padding rows of length $p^{\rightarrow}$ to attach.
- $b^{\rightarrow} = \lceil \lg(c^{\rightarrow}) \rceil$, the number of bits in each width-counter value.
- $h = 16b^{\rightarrow}$ the height of the width-counter block, this accounts for the $b^{\rightarrow}$ bit gadgets composing each counter value
- $c^{\uparrow} = \left\lfloor \frac{n-h}{16} \right\rfloor$, the number of counter values to be counted by a height-counter
- $p^{\uparrow} = n - (h + 16c^{\rightarrow})$, the extra height, or padding, added to the topmost counter value of each height-counter to ensure that the width of the square is exactly $n$. The space and tile type requirements for this are similar to those for $p^{\rightarrow}$.
- $k^{\uparrow} = 2^{\lceil \lg(c^{\rightarrow}) \rceil} - c^{\rightarrow}$, the starting counter value for each height-counter. The representation of this value in the assembly is analogous to that of $k^{\rightarrow}$.
- $k^{\uparrow}_{Max} = 2^{\lceil \lg(c^{\rightarrow}) \rceil} - 1$, the maximum counter value reached by each height-counter (consisting of all 1's in binary). The representation of this value in the assembly is analogous to that of $k^{\rightarrow}_{Max}$.
- $b^{\uparrow} = \lceil \lg(c^{\rightarrow}) \rceil$, the number of bits in each height-counter value
- $w^{\uparrow} = 16 \cdot 2^{\lceil \lg(b^{\uparrow}) \rceil}$, the width of a padded-height-counter, which is a height-counter plus the necessary padding on its right side to fill in the gap between it and the height-counter to its immediate right (must be a power of two multiplied by 16 to allow the width-counter to provide regular binding sites on its north side for padded-height-counters to attach to)
- $pad = w^{\uparrow} - 16b^{\uparrow}$, the width of the padding on the right side of each height-counter (except for the rightmost, which is a special case). This width is $16 \cdot 2^{\lceil \lg \lceil \lg \lfloor \frac{n-h}{16} \rfloor \rceil \rceil} - 16b^{\uparrow}$, which is $O(\lg(n))$, and therefore the number of tile types required to make rows of that length are $O(\lg(n))$
- $\left\lfloor \frac{n}{w^{\uparrow}} \right\rfloor$, the number of padded-height-counters which attach to the north side of the width-counter to form the height-counters block

- $pad^{\dashv} = n \mod w^{\uparrow}$, the width of the extra padding on the right side of the rightmost height-counter. This quantity is bounded by $2w^{\uparrow} - 16b^{\uparrow} - 1$ $\left(\text{or } 16 \cdot 2^{\lceil \lg \lceil \lg \lfloor \frac{n-h}{16} \rfloor \rceil \rceil} - 16b^{\uparrow} - 1\right)$, and therefore by $O(\lg(n))$, and thus the number of tile types required to form rows of this length are $O(\lg(n))$.

- $m = \lg(w^{\uparrow}) = \lceil \lg \left(b^{\uparrow}\right) \rceil$, the number of low-order bits to be matched in the width-counter values to determine the locations to initiate height-counters

- $r = k^{\rightarrow} + \left( w^{\uparrow} \frac{\left\lfloor \frac{n}{w^{\uparrow}} \right\rfloor - 1}{16} \right)$, the rightmost counter value of the width-counter which initiates a height-counter (this height-counter is a special case)

Note that no single component of the construction requires more than $O(\lg(n))$ tile types, whence the tile complexity of the entire construction is $O(\lg(n))$.

### B.1 Hairpin Gadgets

The *hairpin gadgets* are the assemblies in the middle two boxes of Figure 6. Each hairpin gadget is composed of two single-tile-wide paths. These paths are denoted by the thick black lines that represent a series of tiles connected by double-strength bonds. Note that only tile edges through which one of these lines pass, or that contain a colored square, have a non-zero strength glue. The red lines represent a double-strength bond that is the single point of connection between the two paths in each hairpin gadget. The blue squares, two on both the north and south edges of each hairpin gadget, represent single-strength bonds that enable hairpin gadgets to connect to each other in a vertical row.

The top box of hairpin gadgets in Figure 6 (in the second box from the top) are used to represent the most significant bits of the counter and are exactly 13 tiles tall. We construct copies of the other hairpin gadgets that are specific to each of the other bit positions and are 16 tiles tall. Bits are represented as "bumps" and "dents" on the east and west sides of the hairpin gadgets, so that vertical combinations of gadgets create patterns of teeth corresponding to binary numbers. The representation of each bit consists of both a bump and a dent (it is precisely this pattern that determines which bit is being represented in a particular location). We purposely design the two paths that snake through each hairpin gadget to extend the distance of at least one tile into each bump of that gadget.

The white hairpin gadgets form into a column that logically increments a binary counter by representing a binary number $x$ on the west side of the column and $x+1$ on the east side. The grey hairpin gadgets simply present the same binary number on both sides. It is clear that a column of white hairpin gadgets can bind to a column of grey hairpin gadgets via their bottom edge in such a way that the resulting structure is stable at temperature-2 and the adjacent edges between them represent the same binary number. Such columns of gadgets could form in two general ways. First, each white column and each grey column could form in its entirety, and then they could bind together at the bottom to form the full column for a counter if and only if they represent matching numbers. The second possibility is that some incomplete portion of one or both columns is formed when they bind together. Clearly at least the bits which are already represented by both columns will need to match. However, it now must be ensured that as the remaining upper potions of the columns form that they are unable to form "junk assemblies" under the rules of the fuzzy temperature model, which in this case would be assemblies which are stable at temperature-2 but where the columns are incompletely formed to partially represent different numbers. Since this could only occur by hairpin gadgets which attach to the north of columns through their southern edges, by placing the red bonds (see Figure 6), which stabilize both halves of each hairpin gadget together, near the top of those gadgets it is ensured that only hairpin gadgets which have formed all of their bumps and dents (and therefore must match neighboring columns) can attach. In this way it is guaranteed that all columns of white and grey hairpin gadgets which combine together represent the same binary numbers and that no junk assemblies can stably form at temperature-2.

The reason for having two types of hairpin gadgets, both the white and the grey, is as follows. If there were only a single type of hairpin gadget, then a column created from them would have to be able to stably bind to other such columns with bonds equal to strength 2 at the top. While the geometry of the bits represented on the sides of the column would have to represent the correct bits for that column as the columns combine to form a counter, there is nothing to enforce that a column grow the entire necessary height and therefore represent all of the needed bits. For example, if the desired number of bits in each column is $b$, a column of $y = b - x$ hairpin gadgets (where $x < b$) could attach to the right side of a column as long as it matched the first $y$ bits of that column. This could clearly lead to a case where the column which binds to the right of this 'too short' column represents a number which is not exactly 2 greater than the full column that is two positions to the left, thus creating a functionally incorrect counter. By forming each column with two types of hairpin gadgets which bind to each other only at the bottom, and each of which provides a single-strength bond allowing that column to combine to others at the top, it is ensured that the columns can only combine if each represents the correct bits - all of the bits - for that position in the counter.

## B.2 Logical Operation of Hairpin Gadgets

Logically, a hairpin gadget can be thought of as behaving as a scaled-up version (with scale-factor 16) of a single tile. The north and south glues act as strength-2 glues, but are combinations of the two strength-1 glues on the respective sides of the tiles labeled with blue squares on each of those edges. Due to the fault-tolerant design of the hairpin gadgets, the only way that an entire hairpin gadget can form as a stable assembly at temperature 2 is for both of the north glues to match each other, and for both of the south glues to match each other. This essentially mimicks two individual strength 2 glues (*i.e.,* the only way that any two sub-assemblies of a hairpin gadget could stably bind together is for both portions to "agree" on the type of hairpin gadget, since we specifically design all of the interior glues in such a way that the tiles bind exclusively to tiles of the exact same type of hairpin gadget). The strength 2 glues along each of the paths through the hairpin gadget are hard coded to form exactly those paths and provide no logical functionality other than connecting the two sides of the scaled, logical tile together.

The geometric design of the $16 \times 16$ hairpin gadget achieves the fault-tolerance of our main result, which is that no "junk" assembly can stabilize at temperature 2, even when temperature 1 growth is allowed. But when viewed as a single logical tile, each hairpin gadget can be thought of as encapsulating several sets of additional functions beyond fault-tolerance. Figure 5 depicts three separate templates for tile sets, and most hairpin gadgets combine the functionality of two of these. White hairpin gadgets perform the combined functionality of the tile sets in Figure 5a and Figure 5b, while the grey hairpin gadgets, which assemble into the width-counter block, perform the combined functionality of the tile sets in Figure 5b and Figure 5c.

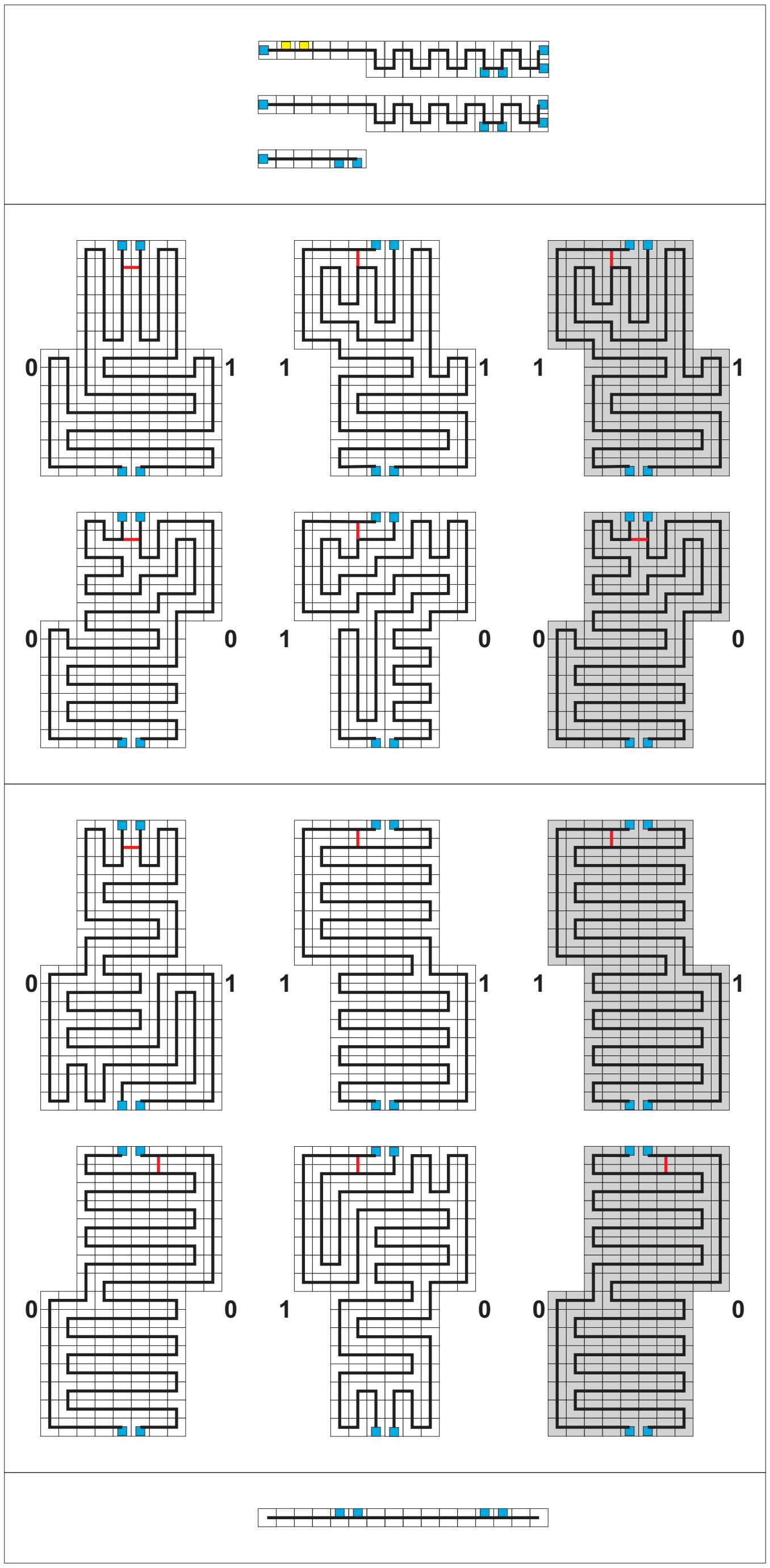

Figure 6: The gadgets that combine to form the counter-values of a counter. The top 6 gadgets that are labeled with bit values are of height 13 rather than 16 for the others, and are used only for the most significant bit in a counter value in order to compensate for the 3 rows of tiles necessary for the gadgets that attach to the top and bottom of the counter and hold the counter values together. Dark black lines represent the strength-2 bonds and forming the bump and dent patterns to represent bit values. The red line is a double bond representing the single point of connection between the two "paths" making up the gadget; see the main text for an explanation of the red bonds' significance. Blue squares represent strength-1 bonds that bind hairpin gadgets to each other and the top/bottom gadgets. Yellow squares represent strength-1 bonds that are used for binding to the vertical counters.

### B.2.1 Increment Tiles

Figure 5a depicts a template for a tile set that is used to form columns that represent pairs of fixed-length binary numbers of the form $x$ (whose bits are represented as the leftmost labels) and $x+1$ (whose bits are represented as the rightmost labels). There are copies of the tile types in the middle group, labeled with an $x$ in the center, specific to each bit position other than the most and least significant positions (the tile types represented in the figure are considered a template for a tile set since copies of each tile type whose center label is "$x$" would have to be generated to be unique to each such bit position). All east and west edges have zero-strength glues, and north and south edges labeled with arrows have double-strength glues that are specific to their bit ordering, allowing tiles that represent bits in each position of significance to bind only to the correct neighbors. The numbers and letters are simply for labeling purposes. It is important to note that this tile set assembles individual columns of tiles, whose labels represent binary values $x$ and $x+1$, *independent* of the order in which the columns come together.

### B.2.2 Tiles to Enforce "$x > k$"

Figure 5b depicts a template for a tile set that assembles columns representing pairs of fixed-length binary numbers of the form $x > k$, where $x$ is a variable binary string whose bits are represented by the labels on left, and $k$ is a fixed value whose bits are represented by the labels on the right. This is merely a template for a tile set since, in order to generate the actual tile types, an input value $k$ must be specified. First, a copy of each of the tile types whose middle label is $i$ must be created specifically for each bit position of $k$ other than the most or least significant bits. Then, all of the tile types whose right labels do not match the bit of $k$ corresponding to the significance of their position are discarded. The resulting tile set consists of tiles that combine (in any order in the two-handed assembly model) to form columns of height equal to the number of bits in $k$, but only in patterns so that the value represented by the string $x$ is greater than $k$. The range of values that $x$ can take on is $k+1, k+2, \ldots, 2^{|k|}-1$ (where $|k|$ is the length of the binary representation of $k$, since $k$ could - and will in our construction - have a 0 as its most significant bit).

### B.2.3 Tiles that Determine Where to Initiate Height-Counters

Figure 5c depicts a template for a tile set that assembles columns representing pairs of fixed-length binary numbers $x$ and $r$, where $x$ is a variable binary string whose bits are represented by the labels on the left, and $r$ is a fixed value whose bits are represented by the values on the right. Additionally, on the north side of the top tile of each column, the glue will specify whether:

1. $x = r$,
2. the low-order $b$ bits of $x$ match those of $r$, or
3. neither of the previous two conditions are true (with the glue for case 1 always being presented if $x$ fully matches $r$, rather than the glue for the second case.

In order to generate this tile set, the input values $r$ and $b$ must be specified. Then, for each of the bit positions $1, 2, \ldots, b-1$, copies of the tile types with the label $i$ are created with glues specific to these bit positions. Next, a copy of each tile type with the label $j$ is created so that the glues position them in the $b^{\text{th}}$-most significant position, and copies are made of the tile types with the label $k$ for each of the remaining positions of greater significance. Finally, tile types whose right labels do not match the bit of $r$ corresponding to the significance of their position are discarded. The resulting tile set consists of tiles that can assemble in any order in the two-handed assembly model to form columns of height equal to the number of bits in $r$. These columns also represent an arbitrary binary string $x$, with the north most glue of the column being "$r$" if $x = r$, "%" if the low-order $b$ bits of $x$ match those of $r$, and "-" otherwise.

#### B.2.4 Combining Tile Set Functionality Into the Hairpin Gadgets

As mentioned previously, most of the hairpin gadgets implement a combination of the functionality of two of the tile sets shown in Figure 5. The white hairpin gadgets perform the combined functionality of assembling into a representation of some binary string $x$ whose value is greater than a given $k$ ($k^{\rightarrow}$ for hairpin gadgets that assemble the width-counter and $k^{\uparrow}$ for those that assemble the height-counters) and less than or equal to the maximum value ( $k^{\rightarrow}_{Max}$ for the width-counter and $k^{\uparrow}_{Max}$ for the height-counters), while representing the value of $x$ on the left and $x+1$ on the right for the width-counter (and bottom and top for the height-counters). The grey hairpin gadgets that assemble into the height-counters perform only the functionality of forming values of $x$ that are within the same ranges as those of their white gadget counterparts. However, the grey hairpin gadgets that assemble into counter values for the width-counter also combine the functionality of presenting glues on their northernmost edges which denote whether $x = r$, whether only the low-order $m$ bits of $x$ match those bits of $r$, or whether neither of these conditions hold. For counter values satisfying one of the first two aforementioned cases, a special top gadget (one of which is shown in the top box of Figure 6 with the two yellow squares on its north side) forms the north piece of that counter value. This is what allows the height-counters to connect in the correct positions.

In order to combine the functionality of two tile sets, tile types from each set are matched up according to the bit positions that they represent. Then, we perform a simple "cross product" by taking each pair of tile types representing the same bit position as well as the same value for $x$ (*i.e.,* they have matching leftmost labels) and making a single, new tile type whose glues and labels contain all of the information from the glues and labels of the two original tile types (with only one copy of the label for $x$).

Finally, for each tile type $t$ in such a generated set, in order to represent $t$ as a hairpin gadget, a final transformation is necessary. For this example we will discuss how to convert a tile type $t$ that (1) represents a bit position of $x$ that is neither the least nor most significant and (2) that combines the "increment $x$" and "enforce $x > k$" functionality with values of 0 and 0 for the bits of $x$ and $x+1$, respectively. Note that transformations for the other classes of tile types is similar and therefore we omit a formal discussion. The form for the hairpin gadget to be constructed is that of the white hairpin gadget in the third box from the top in Figure 6 with the label "$0, 0$."

First, a tile type is created for the position at the southern end of the left path, containing a blue square. Its south glue is strength-2 and contains both arrows from the south glue of $t$ (one of each coming from the tile sets to increment and compare), along with an "L" (since it will be part of the left path through the hairpin gadget), as well as a number representing the significance of the bit. The east and west sides have null glues; the glue for the north side is strength-2, and contains both of the necessary arrows plus the information that both bits are 0, the number representing the significance of the bit, and an 'L1' since it is the first tile on the left path. Next, we add a tile type for each of the interior positions in the left path through that hairpin gadget, with glues that match the north glue of that first tile type, but with each position incrementing the value with "L" so that each tile type is specific to its position in the path and can only attach to its neighbors along that path (also accounting for the direction of the path through each tile type and putting the glues on the appropriate edges). Next, a tile type for the north end of the path is created, analogous to that for the south tile. Then, in a similar matter, tile types for the right path are created (changing the "L" to an "R"). Finally, for the tile types that represent the positions on the two paths where the red bond is located, a strength-2 glue is placed that is unique to exactly those two tile types.

In this manner, all of the tile types required for the hairpin gadgets of the width-counter and height-counters (whose tile types require a 90 degree counter-clockwise rotation from the corresponding tile types of the width-counter) can be generated for a tile set that will assemble into an $n \times n$ square. Notice that the information about the bits of $x$ contained in the labels of each $t$ are now converted to patterns of binary teeth that are used to enforce the correct order of assembly of counter sub-assemblies via geometric constraints.

### B.3 Formation of a Bit Gadget

A *bit gadget* is a logical assembly consisting of a white hairpin gadget and its complementary grey hairpin gadget located on its east side so that all 16 tiles along the adjacent edges of each gadget are in contact with each other. Note that there are only zero-strength glues along this edges so the two hairpin gadgets do not bind to each other. Only hairpin gadgets that encode the same bit along their shared edges can form a bit gadget due to the geometric constraints imposed by the bumps and dents. For the sake of simplicity, hairpin gadgets are designed so that bit gadgets are effectively $16 \times 16$ squares (counting bumps on only one side).

### B.4 Formation of a Counter-Value

There are two types of counters in our construction, the width-counter and a set of height-counters. Both types of counters are fixed-width counters. The width-counter counts from the value $k^{\rightarrow}$ to $k^{\rightarrow}_{Max}$, with each value represented by a binary string of $b^{\rightarrow}$ bits. The height-counters count from the value $k^{\uparrow}$ to $k^{\uparrow}_{Max}$, with each value represented by a binary string of $b^{\uparrow}$ bits. To speak generically of either type of counter, we will use $b$ to denote the number of bits in the values represented by a particular counter. A *counter-value* is an assembly composed of exactly $b$ bit gadgets, along with the top and bottom gadgets that are shown in the top and bottom boxes of Figure 6 (a counter-value is essentially a column). The design of each of the components that make up a counter-value are such that no matter the order in which they assemble, a full counter-value assembly can form only if a valid bit string within the range specified for the counter forms correctly.

### B.5 Formation of the counters

The width-counter is composed of $c^{\rightarrow} - 2$ counter-value sub-assemblies, along with a hard-coded assembly on the left side, which acts as a counter-value representing $k^{\rightarrow}$ via a smooth left side (containing no bumps or dents), and a hard-coded assembly on the right side representing $k^{\rightarrow}_{Max}$ via a smooth right side and rows of length $p^{\rightarrow}$ padding attached to the right. The height-counters are similar but rotated 90 degrees counterclockwise, with $c^{\rightarrow} - 2$ counter-value sub-assemblies, capped by values $k^{\uparrow}$ and $k^{\uparrow}_{Max}$ and northern padding of height $p^{\uparrow}$. Additionally, the height-counters have rows of length *pad* attached to their right sides (except for the rightmost such counter, whose padding rows are of length $pad^{\dashv}$). Figure 7 shows an example sub-assembly of the width-counter.

Only a fully and correctly formed counter-value (*i.e.,* one within the correct range and with the correct shape) can stably bind to another counter-value with the correct bit pattern due to the nature of the two types of top gadgets along with the geometry imposed by the bumps and dents representing bits. In this way, we ensure that counter values can connect to each other only in the correct order and also without "skipping" any columns of the counter. Therefore, the counters all grow to exactly the correct lengths and, moreover, no "junk" assemblies are allowed to stabilize at temperature-2.

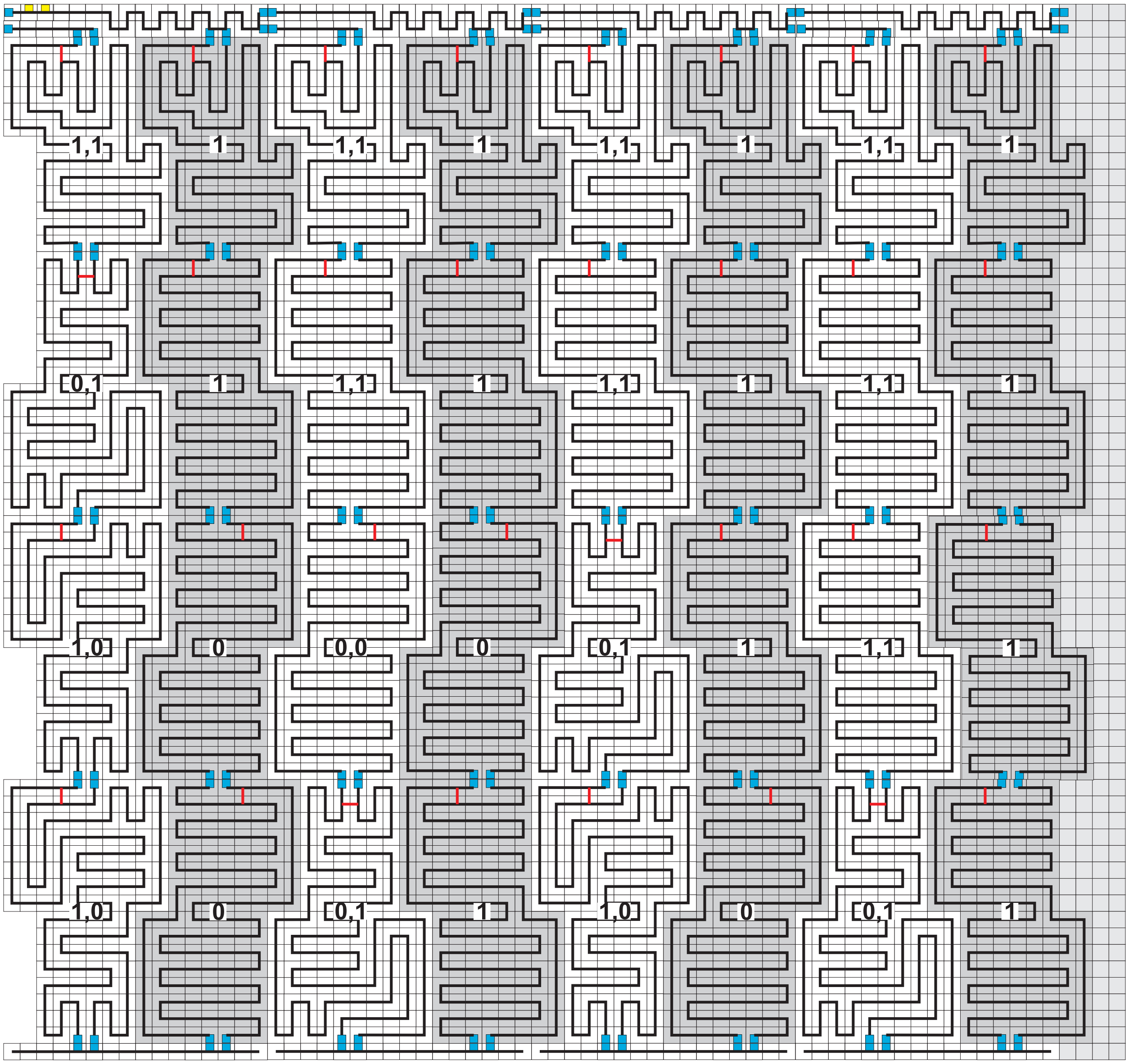

Figure 7: An example of the rightmost 4 columns of the width-counter, counting from 1100 to 1111.

### B.6 Formation of the square

Although the components could actually come together in a number of possible orderings, it is valid to consider just one of them for simplicity of discussion due to the fault tolerant design of the tile set. Therefore, assuming that the width-counter and each of the height-counters correctly and completely form first, then the east side padding for the width-counter, consisting of rows of length $p^{\rightarrow}$, attach and the north side padding for each of the height-counters, consisting of columns of length $p^{\uparrow}$, attach. (Figure 8 shows an example of padding rows attaching.) Next, the east side padding for each of the height-counters, consisting of rows of length $pad$ (or $pad^{\dashv}$ for the special case height-counter that attaches as the rightmost height-counter) attach to each height-counter. Finally, the height-counters attach to the northern edge of the width-counter at the correct locations. The resulting assembly is exactly an $n \times n$ square.

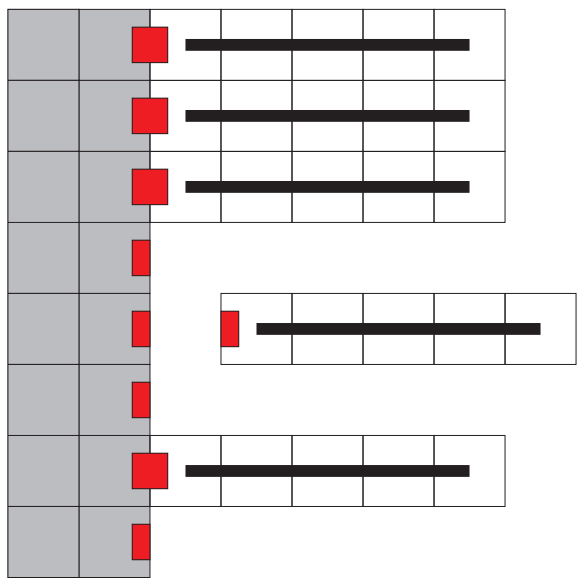

Figure 8: An example of padding rows. The grey portion represents the right side of an assembly to which padding must attach. The padding in this figure is of width 5 and depicted by the rows of white tiles, which are formed by 5 tile types that bind in exactly such a row that binds only on its left side to the correct sub-assembly. In this figure, 4 padding rows have attached and a fifth is nearly in position to do so.